\documentclass[prb,aps,showpacs,twocolumn,preprintnumbers,amsmath,amssymb,superscriptaddress,longbibliography,nofootinbib]{revtex4-2}
\usepackage[english]{babel}
\usepackage{amsmath,amssymb,bbm,mathrsfs,mathtools,multirow,diagbox,xcolor,graphicx,tabularx,comment,amsfonts,dsfont,lettrine,units,comment}

\usepackage{graphicx}
\usepackage[colorlinks=True,linkcolor=blue,citecolor=blue,urlcolor=blue]{hyperref}
\usepackage{bookmark}
\usepackage{xcolor}
\usepackage{chngcntr}
\usepackage{braket}
\usepackage{nicefrac}
\usepackage{bm}
\usepackage{bbm}
\usepackage{braket}
\usepackage{lineno}
\usepackage{array}
\newcolumntype{C}{>{$}c<{$}}
\usepackage{newtxtext} 
\usepackage[normalem]{ulem}

\begin{document}
\bibliographystyle{apsrev4-1}
\title{Four-layer charge density waves and chirality in CsV$_3$Sb$_5$}

\author{Fernando de Juan}
\affiliation{Donostia International Physics Center, 20018 Donostia-San Sebastian, Spain}
\affiliation{IKERBASQUE, Basque Foundation for Science, 48013 Bilbao, Spain}
\author{Mark H. Fischer}
\affiliation{Department of Physics, University of Zurich, 8057 Zurich, Switzerland}

\date{\today}
%%%%%%%%%%%%%%%%%%%%%%%%%%%%%%%%%%%%%%%%%%%%%%%%%%%%%%%%%%%%%%%%%%%%%%%%%%%%%
%

\begin{abstract}
The kagome superconductor CsV$_3$Sb$_5$ is the only one in the AV$_3$Sb$_5$ family (A=K,Rb,Cs) that shows a $2\times2\times4$ charge-density-wave (CDW) ground state competing with the more common $2\times2\times2$. In addition, it is also the only one that shows second-harmonic transport and thus broken inversion symmetry, suggesting these two features are connected. In this work, we test whether the $2\times2\times4$ CDW can break inversion symmetry by analyzing its stacking energetics with a real-space Ginzburg-Landau free energy. In the limit where each layer is forced to adopt a fixed, threefold symmetric tri-hexagonal configuration, we find an analytical phase diagram mostly occupied by inversion preserving AB and ABCD stacked solutions. Relaxing this rigid layer constraint in a physically meaningful way consistent with \emph{ab initio}--calculated parameters, we find several new phases including an AABC solution and a distorted ABCD solution, which break inversion and all mirrors and are thus chiral. However, these phases occupy only small fractions of phase space, suggesting other mechanisms for inversion symmetry breaking may be at play  in CsV$_3$Sb$_5$. 
\end{abstract}
\maketitle

\section{Introduction}
The AV$_3$Sb$_5$ (A=K,Rb,Cs) kagome metals---compounds with V atoms forming layers of kagome structures--- display a remarkable combination of  complex charge and superconducting orders~\cite{ortiz_2019, wilson_2024,Deng24, holbaek_2025}. All three compounds undergo a charge density wave (CDW) transition at $T_{\rm CDW} \sim 100$ K, which takes the form of a $2\times 2$ reconstruction in the kagome plane. This charge order strongly reshapes the electronic structure, impacting  transport, spectroscopy and thermodynamics. However, among the three compounds (abbreviated as KVS, RVS and CVS) the response of CVS is unexpectedly rich, displaying a plethora of additional phenomena, most of which set in at a lower temperature $T^* \sim 30$ K~\cite{wilson_2024}. How these phenomena are related to the charge order is of much current debate.

The CDW ordering, discussed both in terms of  electronic~\cite{Park21, denner_2021a, lin_2021} and lattice instabilities~\cite{Tan21, Christensen21,mcguinness:2025tmp, wang_2026}, 
is attributed to bond modulations with momenta corresponding to the three $M$ points in the Brillouin zone (BZ), which can either form a star of David (SoD) or its inverse, also known as tri-hexagonal (tri-hex) ordering. Interestingly, despite the isoelectronic nature of the three compounds the CDW appears to show significant differences in CVS. First, while for KVS and RVS, the in-plane distortion is believed to always be of tri-hex type, for CVS several groups interpreted their experiments as to require a combination of tri-hex and SoD ordering~\cite{Ortiz21, hu_2022, kang_2023, Kautzsch23, feng_2023}. With respect to the stacking order of the CDW, KVS and RVS show a $2\times2\times2$ distortion, in other words a two-layer periodicity along the $c$ axis with a $\pi$ shift between neighboring planes. In contrast, for CVS several groups report more complicated stackings~\cite{Ortiz21, Wu22u1, Kautzsch23, Stahl22, Xiao23, Plumb24, Sadrollahi25, Wang25, xiao_2023b, Jin24}, with a $2\times2\times4$ order at least at intermediate temperatures~\cite{Kautzsch23, Stahl22, Jin24}, which is quickly suppressed by doping~\cite{xiao_2023b}. Finally, in CVS the whole $ML$ line in the BZ softens all together, instead of a singular phonon wave vector. Recent DFT studies indeed suggest that CVS might be most stable for a $2\times2\times4$ order~\cite{alkorta:2025tmp,chen_2025a}, but many layered orderings appear in close energetic competition, making definitive predictions challenging. 

The symmetry breaking induced by the CDW order leads to complex phenomenology which remains to be fully understood. The reported evidence of time-reversal-symmetry breaking in the charge-ordered phase~\cite{khasanov:2022,Xu22,Liege24,gui_2025} is often interpreted as a consequence of spontaneous loop currents which appear in addition to the lattice distortion~\cite{Fernandes25}. Still, whether such order is established or just on the verge of condensing~\cite{Guo24Tipping} remains under active debate, particularly given the negative results of recent Kerr measurements~\cite{Saykin23, saykin_2025} and the absence of histeresis in the anomalous Hall effect~\cite{yang20,yu_2021,Zheng23,Wang23Rb,Wei24,gan_2021, zhou_2022}. Threefold rotation and vertical mirror symmetry are also reported to be broken in STM~\cite{Wang21, Nie22} and optics~\cite{Xu22}, and even shown to be tunable with $B_z$ and laser polarization in some~\cite{Jiang21, Shumiya21, Deng24,Xing24} but not all~\cite{Li22KSV, Candelora24} STM experiments. These effects are often loosely referred to as chiral, but none of them require the breaking of inversion symmetry.  

In addition to the above, CVS exhibits several unique low temperature phenomena which include field-induced second harmonic transport, also known as magnetochiral anisotropy~\cite{Guo22} (absent in KVS~\cite{Guo24}), field-induced rotation symmetry breaking ~\cite{Guo24Tipping}, many-body interference in magnetotransport~\cite{guo_2025} and hysteretic magnetic susceptibility anisotropy~\cite{gui_2025}. Among these, second harmonic transport is particularly surprising as it is a true bulk signature of inversion symmetry breaking in the material. If both inversion and mirror symmetries are broken, the resulting chiral structure would be consistent with recent XPD and CD ARPES experiments~\cite{Elmers25}. But the connection of these phenomena with charge order is still not understood. With the accepted CDW order parameters, neither the horizontal mirror $z\mapsto -z$ nor inversion can be broken in a $2\times2\times2$ structure. And the potential magnetic order, field-induced or truly condensed, is also inversion-preserving. This suggests the main hypothesis of this work, that the $2\times2\times4$ distortion is responsible for the breaking of inversion symmetry~\cite{Shao24,chen_2025a}, potentially explaning the second harmonic experiments.  

Here, we address the question whether a chiral CDW order is possible within the framework of the charge order discussed so far. For this purpose, we analyze $2\times2\times4$ charge-ordered states within a Ginzburg-Landau free-energy framework. Given the large number of coupling terms in such a description, we consider a mixed representation with the $z$ coordinate, denoting the direction perpendicular to the kagome planes, treated in real space. Thus, the problem is described in terms of layer order parameters akin to a Lawrence-Doniach model for layered superconductors~\cite{lawrence:1970}. This approach has been used to understand the stacking sequence of other CDWs ~\cite{Moncton76,Walker83,Nakanishi84}, and it has the advantage that it allows a more natural interpretation of the hierarchy of couplings in the GL free energy if the system is quasi two dimensional and interlayer couplings are expected to decay fast with distance. Using coupling parameters motivated by recent DFT calculations, we map out the phase diagram of states with four-layer periodicity. We indeed identify several states with chiral point groups (breaking all mirrors and inversion). However, we find these states occupy only a small fraction of phase space. 

The rest of this paper is organized as follows. In Sec.~\ref{sec:symmetry}, we discuss the symmetry of $2\times2\times2$ and $2\times2\times4$ states and why only the latter allow for breaking of inversion symmetry. In Sec.~\ref{sec:landau}, we introduce our approach starting from a Ginzburg-Landau free energy of $M$, $L$, and $U$ order parameters. We then present our main results in Sec.~\ref{sec:results}, the phase diagram of possible phases and their symmetry breaking. We finish with a discussion and conclusions in Sec.~\ref{sec:discussion}. 

\section{CDW Order parameter symmetry}\label{sec:symmetry}
The high-temperature crystal structure of AV$_3$Sb$_5$ has space group 191 (P6/mmm) with point group $D_{6h}$, generated by threefold rotation $C_{3z}$, twofold rotation $C_{2z}$, vertical mirror $\sigma_v$ and horizontal mirror $\sigma_h$. The CDW comprises atomic displacements and electronic modulation mostly within the kagome layers of $V$ atoms. In the single layer limit, the $2\times 2$ CDW order parameter has $M_1^+$ symmetry, with three components $M_n$ representing the modulations for each of the three $\vec M$ points, where $\vec M_1 = (\tfrac{1}{2},0,0)$ and $\vec M_2$ and $\vec M_3$ are obtained by $2\pi/3$ rotations. The isotropic CDW distortions with $|M_1|=|M_2|=|M_3|$ corresponds to the tri-hexagonal ordering when $M_1 M_2 M_3 >0$ and star of David when $M_1 M_2 M_3 <0$. 

In the bulk, K and Rb compounds show a $2\times2\times2$ distortion, which is described in terms of order parameters of $M_1^+$ and $L_2^-$ symmetry~\cite{kang_2023, Kautzsch23, Frassineti23, deng:2025tmp}, The $L_2^-$ distortion is the analog of $M_1^+$ but modulated in the $z$ direction with $q_z=\pi/c$ [$\vec L_1= (\tfrac{1}{2},0,\tfrac{1}{2})$]. The reference frame has the origin at the A atom (between kagome planes), which makes some of the symmetry operations like $\sigma_h$ map between layers. Choosing the kagome plane mirror $\sigma_h'= \tau_c \sigma_h$ as the generator, with $\tau_c$ the translation by one layer, it can be seen that $M_1^+$ and $L_2^-$ transform identically under all point group generators and only differ by their transformation under $t_c$. 

Similarly, the Cs compound shows both $2\times2\times2$ and $2\times2\times4$ order, which can be described in terms of order parameters $M_1^+$, $L_2^-$, and $U_1$ with $\vec U_1 = (\tfrac{1}{2},0,\tfrac{1}{4})$. Again, $U_1$ is the analog of the $M_1^+$ distortion modulated with $q_z = \pm \pi/(2c)$. Note that while $M_n$ and $L_n$ are real-valued order parameters, $U_n$ is in general complex. $U_n$ transforms in the same way as $M_1^+$ and $L_2^-$ under the point group generators except for the fact that $\sigma'_h U_n = U_n^*$. The action of all symmetry generators is listed in Appendix \ref{appA}. 

Given the fact that $L_n$ and $U_n$ are just modulated versions of $M_n$, this combination of order parameters may also be described in an equivalent way with a layered $M_1^+$ order parameter $m_n^l$ (generally a reducible representation), where $n=1,2,3$ and $l=1,\ldots,N$ for $N$ layers, as there is a one-to-one mapping between $M_n$, $L_n$, $U_n$, $U_n^*$ and $m_n^l$ for $N=4$ (or between $M_n$, $L_n$ for $N=2$). This description is useful because of the weak interlayer coupling, and it allows to describe pictorially the different bulk states as stacks of single-layer order parameters. A useful convention to label the different states is to assume a finite value for all $|m_n^l|$ components and classify them by their signs. There are four possible degenerate tri-hex states with $m_1^lm_2^lm_3^l>0$ for a givel layer $l$ centered at the four different origins of the $2 \times 2$ unit cell, which we will label as A,B,C,D. Similarly, the four SoD states with $m_1^lm_2^lm_3^l<0$ will be labeled as a,b,c,d. In this labeling convention, a state with only $M_n$ order parameters, often called the $MMM$ state, is thus described as the AA state for two layers, or the AAAA state for four layers. For simplicity we will refer to this state with its minimal periodicity and call it the A state. Other examples can be described in this way like the $LLL$ state, an alternating tri-hex / SoD structure of the type Aa, or the $MLL$ state with $M_n=(M,0,0)$ and $L_n=(0,L,L)$, which is labeled as the AB state. Note this is just a labeling convention and makes no reference to the actual symmetry of the state, which depends on the values $m_n^l$

Regardless of the choice of representation of order parameters, their symmetry analysis already reveals two important general features of the CDW order. First, all order parameters consisdered are even under the $z$-axis twofold rotation $C_{2z}$, which is therefore never broken within this theory. Since $D_{6h} = D_{3h} \times C_{2z}$, the symmetry-breaking pattern can be understood most easily in terms of $D_{3h}$ only, which is generated by $\{C_3,\sigma_v,\sigma'_h\}$. Second, the horizontal intralayer mirror symmetry $\sigma'_h$ cannot be broken for any $2 \times 2\times2$ CDW. It is only when more than two layers are considered that this symmetry can be broken. The same applies to the intralayer inversion symmetry $I = C_{2z}\sigma'_h$. Since a chiral state requires breaking of $I$ and $\sigma'_h$, no chiral state can be realized within two-layer CDWs. Crucially, this remains even true when a loop-current (or flux) order is included~\cite{Fernandes25}, which is believed to have symmetry $mM_2^+$, where $m$ denotes a magnetic or time-reversal odd order, as the combined order then can couple to a $z$-axis magnetic field~\cite{Christensen22, Wagner23}. However, an open question regards the possibility of additional symmetry breaking at $T^*$ with new order parameters, which could break further symmetries including inversion~\cite{tazai_2025,Ingham25,Zhang24B1u}.

The purpose of this work is to explore whether chiral charge order can be realized at all in the context of a Ginzburg-Landau analysis for four layer CDWs. In the following, we will present such analysis and argue why the layered represenation is more suited to capture the relevant physics with a controllable set of parameters. 

\section{Landau free energies}\label{sec:landau}
The general Landau free energy for the $M$, $L$ and $U$ order parameters, constrained by symmetry with the previous representation, takes the form \cite{Christensen21, Park21, Wagner23}
\begin{align}\label{eq:fullF}
F = F_M + F_L + F_U  +F_{ML} + F_{MU} + F_{LU}.
\end{align}
Here, $F_M$, $F_{L}$, and $F_{U}$ are, up to fourth-order terms, given by
\begin{align}
    F_M &= a_M M^2 + b_M M_1 M_2 M_3 + c_M^{(1)} M^4 \nonumber\\
    & + c_M^{(2)} \sum_n M_n^2M_{n+1}^2,\label{eq:FM}\\
    F_L &= a_L L^2 + c_L^{(1)} L^4 + c_L^{(2)} \sum_n L_n^2L_{n+1}^2,\label{eq:GLL}\\
    F_U &= a_U U^2 + c_U^{(1)} U^4+c_U^{(2)} \sum_n |U_n|^4 + 2c_U^{(3)} \sum_n{\rm Re} U_n^4\nonumber \\
    &+ 2c_U^{(4)} \sum_n{\rm Re} U_n^2 U_{n+1}^2 + 2c_U^{(5)} \sum_i{\rm Re}U_n^2 U_{n+1}^{*2},
\end{align}
and contain only the respective order parameters. Further, the index $n$ is periodic modulo 3, i.e. $M_{n+3}=M_n$, $M^2 = \sum_n M_n^2$ and the same applies for $L_n$ and $U_n$. The terms describing mixing of different order parameters are given by
\begin{align}
    F_{ML} &= b_{ML} \sum_n M_n L_{n+1} L_{n+2} + + c_{ML}^{(1)} M^2 L^2 \nonumber\\&+ c_{ML}^{(2)} \sum_n M_nM_{n+1}L_n L_{n+1}  + c_{ML}^{(3)}  \sum_n M_n^2 L_n^2, \label{eq:FML}\\
    F_{MU} &= 2 b_{MU} \sum_n {\rm Re}[U_n U_{n+1}^*]M_{n+2} + \ldots\\
F_{LU} &=2b_{LU} \sum_n{\rm Re}[U_n U_{n+1}]L_{n+2} +\ldots  \label{eq:FLU} 
\end{align}
Up to fourth order, this free energy has twenty-eight independent parameters---three for the quadratic, four for the cubic, and twenty-one for the quartic terms---an unfeasibly large number.

This free energy can be reduced to the two-layer case simply by setting $U_n=0$. For this case, the values of the coefficients were computed ab-initio at zero temperature in Ref. \cite{Ritz23}, finding $a_M \approx a_L<0$ (below the transition) and $b_M \approx b_{ML} <0$. While both cubic terms being negative would seem to favor an $Aa$  state ($MMM+LLL$)~\cite{Christensen21}, the dominant quartic couplings were found to be $c_{ML}^{(2)}>0$ and $c_{ML}^{(3)}>0$, which are minimized by the $AB$ ($MLL$) or $ab$ ($\tilde{M}LL$) states instead. Finally, with $b_{ML}<0$, the ground state is $AB$, which is the one believed to be realized in KVS and RVS. Whether there is a single transition to $AB$, or two transitions from $Aa$ to $AB$ depends on the temperature evolution of the quadratic coefficients, which is not predicted ab-initio.   

\begin{figure}[t]
    \includegraphics{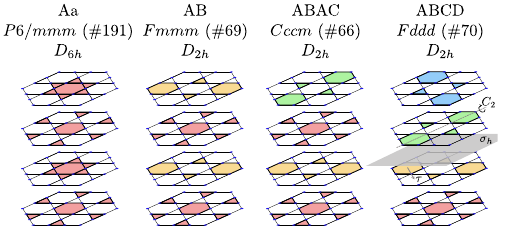}
     \caption{Charge-density-wave ground states in AV$_3$Sb$_5$ within the rigid-layer limit and their symmetry. a) The Aa alternating tri-hex-SoD (Aa) or LLL state. b) The AB state or MLL state, the ground state of the A$=$K,Rb compounds. c) The ABAC state and d) The ABCD state. For the latter, the in-plane $C_2$ axis, as well as the glide mirror $\sigma_h$, a mirror combined with a translation by $\tau$, are marked to emphasize the $D_{2h}$ point group symmetry.}
     \label{fig:rigidlayer}
\end{figure}

\subsection{Free energy in layer representation}

The large increase in the number of parameters when going from two to four layers in the $MLU$ representation makes it unfeasible to explore the phase diagram simply by mapping the values of the coefficients. A possible strategy could be to employ perturbation theory to extend the $ML$ free energy by considering small coefficients for all terms involving $U$. However, this strategy would contradict the general expectation that the layers are weakly coupled: a given term in the $MLU$ free energy maps to many terms in the layer free energy involving arbitrary neighbors, and the constraint that this term is small maps into a constraint between many terms, which may separately be large while satisfying the constraint. Hence, a better suited approach is to first map the $ML$ free energy to a layer free energy exactly, and then consider extending this free energy by considering the real space coefficients as perturbations. Here, the assumption is that couplings remain short ranged, but $U$ need not be small. In this way, the physics of nearly independent layers can be kept, and one can ask whether controlled departures from this theory allow for any chiral CDW states.  

Given this logic, we will proceed in the following in three steps: (1) we change form $MLU$ to a layer representation, keeping only nearest-layer contributions for the fourth-order terms. (2) To get an understanding of the general phase diagram, we start from a \emph{rigid-layer construction}, where all components have the same absolute value. (3) For numerical calculations away from the rigid-layer limit, we minimize the free energy with parameters calculated in Appendix \ref{app:coeffs}.

The layer free energy is expressed in terms of the independent $M$ vectors for each layer $l$, which we denote by $m_n^l$. As before, the $n$ index is periodic $m^{l}_{n+3} \equiv m^l_n$. The free energy per layer 
can be written for an arbitrary number of layers as  
\begin{align}
F_N = \frac{1}{N} \left( F^{(2)} + F^{(3)} + F^{(4)}   \right),
\end{align}
for $N$ layers, where $F^{(n)}$ contains terms in $n$-th order. In the main text, we will employ periodic boundary conditions in a four layer unit cell so that $m^{l+4}_{n} \equiv m^l_n$. The allowed quadratic terms of the free energy are
\begin{align}
F^{(2)} = \sum_{n,l} a_1 (m_n^{l})^2 + a_2 m_n^lm_{n}^{l+1} + a_3 m^l_n m^{l+2}_{n}.
\end{align}

Similarly, the allowed third-order terms are
\begin{align}
    F^{(3)} &= \sum_{l}\Big\{b_1 m_1^{l}m_{2}^{l}m_{3}^{l} \nonumber \\ 
    &+  \sum_n\big[b_2(m^l_{n}m^{l}_{n+1}m^{l+1}_{n+2}+m^l_{n}m_{n+1}^{l+1}m_{n+2}^{l+1}) \nonumber\\&+ b_3 (m_n^{l}m_{n+1}^{l}m_{n+2}^{l+2}+m_n^{l+2}m_{n+1}^{l+2}m_{n+2}^{l}) \nonumber\\ &+ b_4( m_n^{l}m_{n+1}^{l+1}m_{n+2}^{l+2}+m_n^{l+2}m_{n+1}^{l+1}m_{n+2}^{l})\big] \Big\}.
\end{align}

For the quartic terms, finally, 21 terms are symmetry allowed. As discussed above, we restrict the quartic terms in the free energy to contain only terms up to first neighbors, yielding

\begin{spreadlines}{10pt}
\begin{align}
&F^{(4)} =  c_1  \sum_l [\sum_n (m_n^i)^2]^2 + \sum_{l,n} \Big(c_2 (m_n^l)^2(m_{n+1}^l)^2 \nonumber\\
+& c_3  (m_n^l)^2 (m_n^{l+1})^2 + c_4  [(m_n^l)^3 m_n^{l+1}+(m_n^{l+1})^3 m_n^l] \nonumber \\ 
+&  c_5  m_n^l m_n^{l+1}\left[(m_{n+1}^{l+1})^2+(m_{n+2}^{l+1})^2+(m_{n+1}^{l})^2+(m_{n+2}^{l})^2\right] \nonumber\\+&c_6 m_n^lm_{n+1}^l m_n^{l+1}m_{n+1}^{l+1} \nonumber\\ 
  +&  c_7 (m_n^l)^2 \left[ (m_{n+1}^{l+1})^2+(m_{n+2}^{l+1})^2 \right]\Big).
\end{align}
\end{spreadlines}
The coefficients for this free energy can be derived from those computed \emph{ab initio} for the $ML$ free energy in Ref. \cite{Ritz23}. We consider this mapping in detail in Appendix ~\ref{app:coeffs}.

\section{Results}\label{sec:results}
\subsection{Rigid-Layer Limit}

To analyze the possible ground states of the free energy, it is instructive to start with the limit where each layer is uncoupled from the rest. This free energy is equivalent to that of the $M_n$ order parameter alone ($F_M$ in Eq. \ref{eq:FM}). It has a single cubic term $b_1$ and two quartic terms $c_1$ and $c_2$. The term $c_1$ only fixes the modulus of the order parameter, and the term $c_2$ chooses between single-Q $(1,0,0)$ and triple-Q $(1,1,1)$ ground states. Experimentally, the ground state is always triple-Q, so we fix $c_2<0$. The coefficient $b_1$ then decides between tri-hex and SoD solutions, so we take $b_1<0$ since tri-hex is preferred~\cite{Tan21,Xiao23}. In this limit, all components of the order parameter take the same value $|m_n^l| = m_0$, a feature which holds approximately in \emph{ab initio} calculations~\cite{alkorta:2025tmp}.

In this limit, one can only have a discrete set of states for each layer given by the four tri-hex states $m_{\rm A}= m_0(1,1,1)$, $m_{\rm B}= m_0(1,-1,-1)$, $m_{\rm C}= m_0(-1,1,-1)$, and $m_{\rm D}= m_0(-1,-1,1)$. As such, a model with arbitrary number of layers can be mapped to a four-states Potts model, discussed extensively for kagome systems in Ref.~\cite{Park21}. If the coefficient $b_1$ is small, one may also consider including the SoD states, which are given by $m_{\rm a} = -m_{\rm A}$ and the same for ${\rm b, c, d}$. While the state might deviate at finite temperatures from such a rigid-layer state, we can imagine the states to be good approximations at $T\rightarrow 0$.

Table~~\ref{tab:rigidlayertable} lists the seven possible states with a four-layer unit cell focusing on tri-hex states only, along with their degeneracies and symmetry groups. The sum of all degeneracies equals $4^4 = 256$, corresponding to four kagome unit cells on four layers. This analysis reveals that in the rigid-layer limit, the only chiral state with four layers is AABC. 
\begin{table}[t]
\begin{center}
\begin{tabular}{|c|c|c|c|} 
 \hline
 State & Deg. & SG & PG \\ 
 \hline
 $AAAA$ & 4 & P6/mmm (\#191) & $D_{6h}$\\
 \hline
 $ABAB$ & 12 & Fmmm (\#69) & $D_{2h}$\\
 \hline
 $AABB$ & 24 & Fmmm (\#69) & $D_{2h}$\\
 \hline
 $ABCD$ & 24 & Fddd (\#70) & $D_{2h}$\\
 \hline
 $AAAB$ & 48 & Cmmm (\# 65) & $D_{2h}$\\
 \hline
 $ABAC$ & 48 & Cccm (\#66) & $D_{2h}$\\
 \hline
  $AABC$ & 96 & C222 (\# 21) & $D_{2}$\\
   \hline
\end{tabular}
\caption{Symmetry and degeneracy of possible states of the rigid-layer model. Note within the rigid-layer approximation, only AABC is chiral with point group $D_2$.}
\label{tab:rigidlayertable}
\end{center}
\end{table}

\begin{figure*}
         \includegraphics[width=0.98\textwidth]{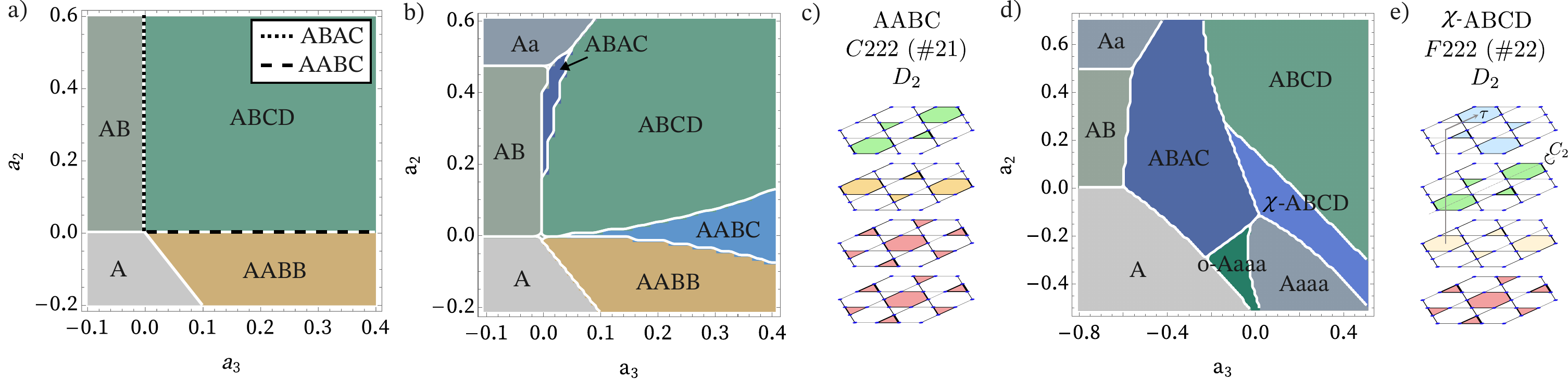}
     \caption{a) Rigid layer phase diagram as a function of quadratic couplings. White lines mark phase boundaries, and dotted/dashed lines denote a phase boundary where a third solution, shown in the inset, is also degenerate. b) Numerical phase diagram with Ritz parameters, where the rigid-layer constraint is relaxed. c) Structure of the chiral solution AABC found in b). d) Numerical phase diagram with modified Ritz parameters with $\tilde{c}_2=0.75 c_2$ and cubic terms with values $(x,y)=(3.1, 3.7)$. e) Structure of chiral ABCD state found in d), which lacks the mirror symmetry of the ABCD state (Fddd). The translation due to F centering is shown as an arrow.}  
     \label{fig:Fig2new}
\end{figure*}

The minimization of the free energy in the rigid-layer limit with only tri-hex states is particularly simple. Using the product rules
\begin{align}
m_{n}^lm_{n+1}^l &= m_0 m_{n+2}^l, \\
(m_{n}^l)^2 &= m_0^2,
\end{align}
the full four-layer free energy reduces to
\begin{align}
F = F_0 &+ \sum_{n,l} A_2 m_i^n m_{l}^{n+1} + A_3 m_l^n m_{l}^{n+2} \nonumber \\&+ b_4( m_n^{l}m_{n+1}^{l+1}m_{n+2}^{l+2}+m_n^{l+2}m_{n+1}^{l+1}m_{n+2}^{l}), 
\end{align}
where
\begin{align}
    A_2 &= a_2+2 b_2m_0+(2c_4+4c_5+c_6)m_0^2,\\
    A_3 &= a_3+2b_3m_0,
\end{align}
and 
\begin{align}
F_0 = N [3 a_1 m_0^2 \!+\! b_1 m_0^3\! +\! 3(c_1 \!+\! 3c_2 \!+\! c_3 \!+\! 2c_7)m_0^4].      
\end{align}
Note that $F_0$ is the part of the free energy that is the same for any stacking, in other words it depends only on $m_0$. So for a given $m_0$, one can write the whole phase diagram in terms of the two dimensionless parameters $A_3/A_2$ and $b_4/A_2$. It is not difficult to show that this would have been the case even when including quartic, second-neighbor terms, since these would at most renormalize the values of $A_2$, $A_3$ and $b_4$.

Figure~\ref{fig:Fig2new}(a) shows the phase diagram of the rigid-layer construction as a function of the quadratic couplings $a_2$ and $a_3$ for the parameters $b_1=-1$ and $c_1=1$. The phase diagram as a function of the dimensionless parameters $A_3/A_2$ and $b_4/A_2$ is shown in the appendix in Fig.~\ref{Fig4}. We find that only four of the rigid-layer phases are stabilized over an extensive region in parameter space, namely the single-layer A state, the two-layer AB state, and the four-layer states with ABCD and AABB with the first state having point group $D_{6h}$ and the others $D_{2h}$, see Tab.~\ref{tab:rigidlayertable}. In contrast, the states AABC and ABAC only occur along the degeneracy lines between AABB and ABCD and AB and ABCD, respectively. Therefore, within the rigid-layer limit there are no chiral ground states. In the next section, we consider the numerical minimization away from the rigid-layer limit and explore whether AABC or any other chiral states may be obtained.

\subsection{Numerical minimization of nearest-neighbor model}

To relax the rigid-layer constraint, we perform an unrestricted minimization of the layer free energy. We compute the cubic and quartic parameters by mapping the nearest neighbor free energy to the $ML$ free energy, for which the parameters were computed in  Ref.~\cite{Ritz23} (see Appendix \ref{app:coeffs}). We consider the quadratic parameters $a_2$ and $a_3$ as free coefficients to draw the phase diagram, which is shown in Fig.~\ref{fig:Fig2new}(b). In general, the phase diagram shows a very similar structure to the rigid-layer one in Fig.~\ref{fig:Fig2new}(a), which shows that the ab initio free energy is in fact quite close to the rigid layer limit. However, three new phases appear as well in the unrestricted phase diagram: First, the phase boundaries, where ABAC and AABC phases appeared, became finite regions in the phase diagram. In addition, the Aa phase appears at large $a_2$, which shows phases including SoD layers are also energetically allowed. The parametrization of all the obtained solutions with their space groups is shown in Appendix \ref{app:parametrization}. Interestingly, all of these phases have been found as ground states in \emph{ab initio} calculations before, most often quasi-degenerate in energy \cite{Xiao23,Shao24,Chen24SCHA,alkorta:2025tmp}.

Our calculation shows that there is in fact a chiral CDW in the phase diagram, the AABC phase, with the parameters computed from Ritz \emph{et al.}~\cite{Ritz23}, consistent with the findings in Ref.~\cite{Shao24}. However, it appears in the rather unphysical region where $a_3 \gg a_2$, in other words, the second-neighbor coupling is much larger than the first. In contrast, in the region with $a_2 > a_3$ the most common phases obtained both in \emph{ab initio} calculations and in X-ray refinements are found: the Aa and AB two-layer states, and the ABAC and ABCD four-layer states. The narrow ABAC sliver between AB and ABCD simply reflects the fact that the dominant first-neighbor couplings prefer any state with local AB stacking, without a strong preference for the relative second-neighbor stacking. 

Therefore, within the approximation of nearest-neighbor Ritz parameters, it appears very unlikely that a chiral phase can exist. Nevertheless, given the experimental transport evidence of broken inversion symmetry, we next consider more general parametrizations that are consistent with the Ritz parameters to widen the search for chiral phases. 

\subsection{Extensions of the nearest-neighbor model}\label{sec:extension}

To proceed, we first relax the assumption that the cubic terms are nearest neighbor only. Since the DFT calculation of the cubic coefficients was done in a two-layer unit cell, there is actually a two parameter manifold of values that is consistent with the DFT result. By imposing two-layer periodicity in the four-layer free energy, the $b_i$ coefficients can be written in terms of two parameters $(x,y)$ as detailed in Appendix \ref{app:coeffs}.
The Ritz parameters are obtained when $x=y=0$. The phase diagram for the Ritz parameters as a function of $x,y$ for $a_2=0$ and $a_3=0$ is shown in Appendix \ref{app:additional}. This phase diagram includes the new phases ABAb, AAAa, and Aaaa. It is worth noting that some X-ray refinements \cite{Wu22u1,Kautzsch23} do find such type of mixed SoD/tri-hex states. However, none of these states are chiral (see Table \ref{tab:beyondrigidlayertable}). 

The same procedure could in principle be applied to the quartic coefficients to explore the parameter space that is consistent with the DFT predictions of the two layer model. Given the large number of parameters, this exploration is, however, unfeasible. With the physical motivation of allowing further relaxation away from the rigid-layer constraint, we consider a minimal extension of the model where we lower the value of the intralayer coefficient $c_2$. We find that a moderate reduction of $\tilde{c}_2 = 0.75c_2$ already allows for new phases. The phase diagram obtained with this set of parameters is shown in Appendix \ref{app:additional} for $a_2=a_3=0$. Interestingly, this minimal change allows two additional phases, an orthorhombic Aaaa, which we denote as o-Aaaa in short, and a chiral ABCD phase, which we denote as $\chi$-ABCD. The chiral ABCD phase occurs around $(x,y)=(3.1,3.7)$. Fixing these values, we can thus calculate a new phase diagram as a function of the quadratic coefficients $a_2$ and $a_3$, which is shown in Fig. \ref{fig:Fig2new}(d). The structure of the chiral ABCD phase is illustrated in Fig. \ref{fig:Fig2new}(e), showing that the horizontal mirror is broken. Note that the values $(x,y)=(3.1,3.7)$ correspond to cubic couplings $b_1=-8.73$, $b_2=3.82$, $b_3=-0.51$ and $b_4 = -3.7$, so one of the second-neighbor couplings is similar in magnitude to the first-neighbor one, $|b_2| \sim |b_4|$. In this sense, this set of parameters is an extension of the first-neighbor model, which is still consistent with the \emph{ab initio} parameters computed in Ref.~\cite{Ritz23}. A summary of all the phases beyond the rigid-layer limit is provided in Table~\ref{tab:beyondrigidlayertable}.

\begin{table}[t]
\begin{center}
\begin{tabular}{|c|c|c|c|} 
 \hline
 State & Deg. & SG & PG \\ 
 \hline
  $Aa$ & 8 & P6/mmm (\# 191) & $D_{6h}$\\
 \hline
 $AAAa$ & 8 & P6/mmm (\# 191) & $D_{6h}$\\
 \hline
$Aaaa$ & 8 & P6/mmm (\# 191) & $D_{6h}$\\
 \hline
  Orth. $Aaaa$ & 48 & Cmmm (\# 65) & $D_{2h}$\\
 \hline
  $ABAb$ & 48 & Cmmm (\# 65) & $D_{2h}$\\
 \hline
  Chiral $ABCD$ & 48 & F222 (\# 22) & $D_{2}$\\
 \hline
\end{tabular}
\caption{Symmetry and degeneracy of the states found numerically beyond the rigid layer ones. Note there are hexagonal and orthorhombic versions of Aaaa. }
\label{tab:beyondrigidlayertable}
\end{center}
\end{table}

\subsection{Symmetry-breaking in the charge order}

\begin{table}[b]
\begin{center}
\begin{tabular}{|c|c|c|c|c|c|c|} 
 \hline
Irrep & $\Gamma_1^+$ & $\Gamma_1^-$& $\Gamma_2^+$& $\Gamma_2^-$ & $\Gamma_5^+$ & $\Gamma_5^-$ \\ 
 \hline
& $A_{1g}$ & $A_{1u}$ & $A_{2g}$ & $A_{2u}$ &  $E_{2g}$ & $E_{2u}$  \\ 
 \hline
$\sigma_v$  & $+$ & $-$& $-$& $+$ &  &  \\ 
 \hline
$\sigma_h$  & $+$ & $-$& $+$& $-$ & $+$ & $-$ \\ 
 \hline
\end{tabular}
\caption{Translation trivial, $C_{2z}$ even irreps of SG 191, relevant for secondary order parameters.}
\label{tab:irreps}
\end{center}
\end{table}

To understand the phenomenological implications of symmetry breaking it is often enough to consider the point group symmetry and disregard translations. An illuminating way to quantify the breaking of symmetries is to compute secondary order parameters that are trivial under translations (i.e. zero momentum or $\Gamma$ irreps) and can only become finite when certain symmetries are broken. Since $C_{2z}$ can never be broken by secondary order parameters (because the primary ones are all even) we list all the $C_{2z}$ even $\Gamma$ irreps of SG 191 along with their parity under the mirror symmetries $\sigma_v$ and $\sigma_h$ in Table \ref{tab:irreps}. 

The breaking of $C_3$ symmetry can be quantified by the  $\Gamma_5^+$ ($E_{2g}$) irrep, which represents a nematic order parameter. Up to nearest neighbors there are two possible such irreps,
\begin{align}
(N_x,N_y) & = \sum_{l,n} (\delta^n_x,\delta^n_y) m_n^l m_n^{l},\\
(N'_x,N'_y) & = \sum_{l,n} (\delta^n_x,\delta^n_y) m_n^l m_n^{l+1} ,
\end{align}
using $(\delta^n_x,\delta^n_y) = (\{1,-1/2,-1/2\},\{0,\sqrt{3}/2,-\sqrt{3}/2\})$. The first measures the average intralayer nematicity, while the second can be considered an interlayer nematicity, related to the one defined in \cite{Park21}. The interlayer nematicity is finite if the stacking pattern globally breaks threefold symmetry even if locally each layer preserves a threefold center. Since the nematic components obtained in the numerical evaluation can vary in direction for point to point in the phase diagram, we report the modulus $N_{\rm intra} = \sqrt{N_x^2+N_y^2}$ and $N_{\rm inter} = \sqrt{N_x'^2+N_y'^2}$. 

\begin{figure}
    \includegraphics[width=0.48\textwidth]{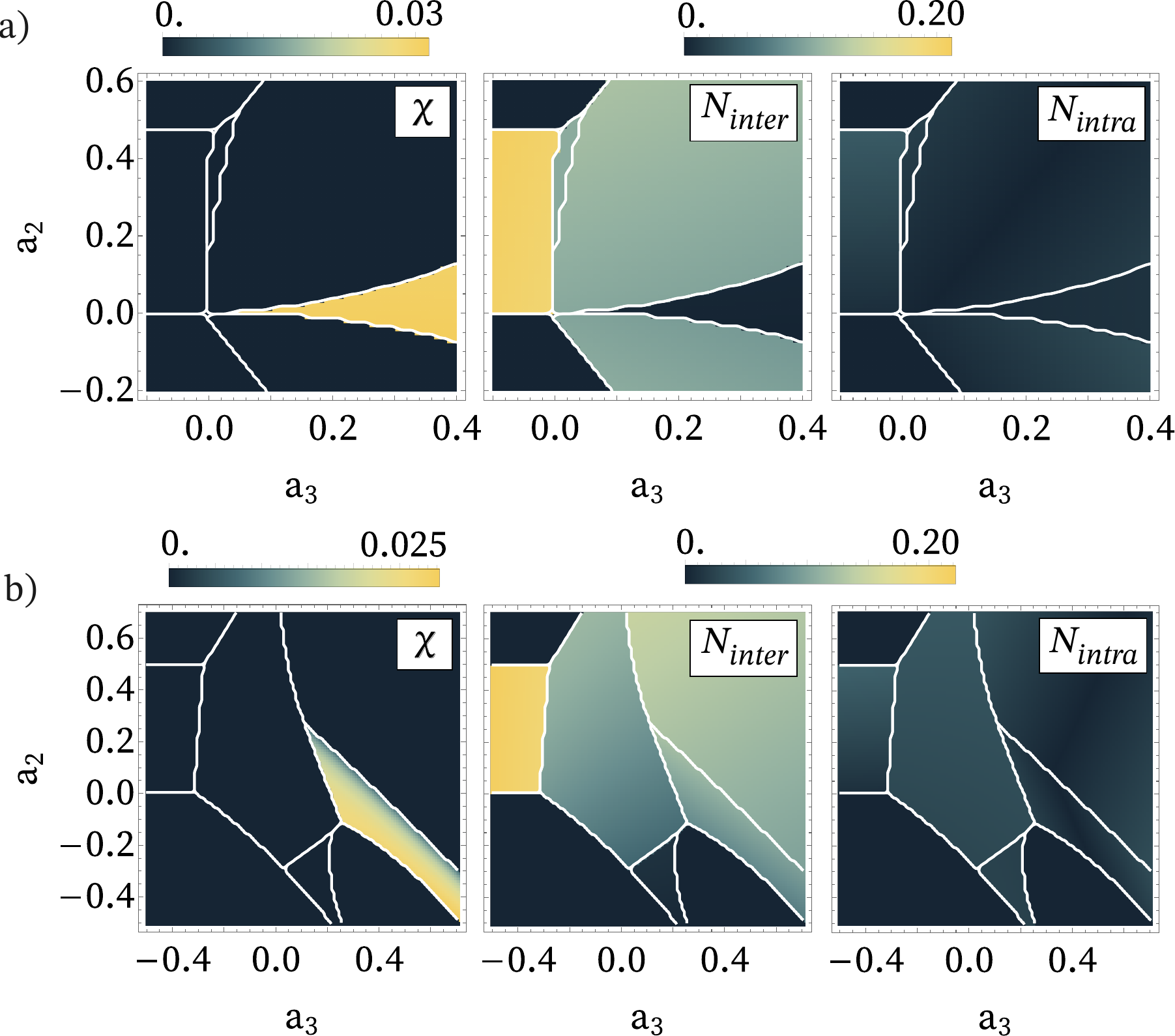}
     \caption{Order parameter maps corresponding to the phase diagrams in Fig. \ref{fig:Fig2new}. Chirality (left), interlayer nematicity (middle) and intralayer nematicity (right) are shown for a) Ritz and b) modified Ritz parameters.}  
     \label{Fig3}
\end{figure}

Chirality, in contrast, can be quantified by a $\Gamma_1^-$ order parameter ($A_{1u}$) which is odd under all mirrors and inversion. As stated in the introduction, such a secondary order parameter must involve at least three different layers, and it can be written as 
\begin{align}
\chi &= \sum_{l,n} m_n^{l} m_{n+1}^{l+1}m_{n+2}^{l+2} - m_n^l m_{n+1}^{l+2}m_{n+2}^{l+1}.
\end{align}
Figure~\ref{Fig3} shows the computed nematic and chiral order parameters for the two sets of parameters used in the phase diagrams of Fig. \ref{fig:Fig2new}. The figure clearly reveals that finite chirality only emerges for the AABC phase region in Fig. \ref{Fig3}(a), and for the chiral ABCD region in Fig. \ref{Fig3}(d). In addition, we observe that the AABC phase shows negligible nematicity despite formally having $D_2$ symmetry, while the chiral ABCD phase, like its non-chiral version, does show significant interlayer nematicity. This can be understood from the rigid-layer limit, where the AABC has three contributions to interlayer nematicity (from AB, BC and CA pairs of layers), which give nematic vectors rotated by 120 degrees so that the overall nematicity adds up to zero. This is not the case for ABCD, where there are four contributions adding up to a finite value. The rigid-layer limit also explains why the AB phase has the strongest interlater nematicity, with two equal contributions that add up. The difference in these two types of nematicities may help understand the breaking of $C_3$ in transport experiments, since the intralayer nematicity, more relevant for transport, is significantly smaller than the interlayer one. 

\section{Discussion}\label{sec:discussion}

In this work, we have found two possible chiral CDW ground states of a physically motivated free energy to describe the $2 \times 2 \times 4$ states of CVS. Both share the $D_2$ point group, which does allow second harmonic transport and could potentially explain the transport experiments as well as other reports of chirality. Importantly, while the observed second harmonic signal only appeared at finite field, zero-field second harmonic transport is already allowed for $D_2$ in the component $\chi_{xyz}$, which is not conventionally measured and could represent an additional check for the existence of chirality. 

However, the states found in this work either require an unnatural hierarchy of quadratic couplings $a_3\gg a_2$ (for AABC) or an extension of the nearest neighbor model derived from Ritz (for $\chi$-ABCD). Given this, one may consider what other sources of symmetry breaking may be responsible for the observed second harmonic transport. The possibility of finite-size effects like surface transport appears to be an unlikely explanation, as in metallic systems transport is dominated by bulk states. Restricting to bulk mechanisms, one possibility is that an additional, inversion-breaking order parameter has so far remained undetected, as in recent theoretical proposals \cite{tazai_2025,Ingham25}.

Another possibility is that loop current order, possibly activated at a lower temperature than the CDW transition, combines with the CDW in such a way as to break inversion symmetry. The most commonly proposed irrep for such flux order is, however, $M_2^+$, which is inversion even, so its trivial addition to any of the non-chiral CDW states is not sufficient to explain the transport experiments. A more complex explanation may be that the presence of flux order significantly restructures the phase diagram and causes a transition to one of the chiral states. It may also be the case that flux order develops a non-trivial stacking beyond the $M_2^+$ irrep. While these proposals may be worth exploring, they are currently beyond the proposed loop current order, and further confirmation of their existence would also be required. A further aspect we have not considered is the possible out-of-plane incommensuration of the CDW. While experimentally the Bragg reflections appear at the commensurate wavevector $U$ within resolution, the $U_1$ irrep formally allows a Lifshitz invariant (linear in the $\partial_z$ derivative), which allows for a first transition to an incommensurate state, which may then lock-in to a commensurate state at lower temperatures~\cite{McMillan76,Bruce78}. 

Finally, as more experiments contribute to scrutinize the symmetries of the CDW ground state in CVS, it is worth emphasizing that the breaking of the vertical mirror symmetries, often observed in STM experiments, is not enough to guarantee a bulk chiral structure. The breaking of the vertical mirrors is possible in two-layer states as well and can also give rise to surface natural circular dichroism \cite{Xu22} with bulk inversion or the horizontal mirror $\sigma_h$ being preserved. Second harmonic transport \cite{Guo22,Guo24}, on the other hand, is more bulk sensitive and represents a stronger evidence of the breaking of inversion symmetry. However, why optical second harmonic generation does not show any features in CVS remains unexplained~\cite{Ortiz21,Xu22}.  

We hope that our work will contribute to the understanding of the complex phase diagram of CVS, providing a framework to elucidate the relationship between $2\times2\times4$ CDWs, nonlinear transport and the breaking of inversion symmetry. 
 
\section{Acknowledgements} We thank Manex Alkorta, Ion Errea, Chunyu (Mark) Guo, Sofie Castro Holb\ae k, Philip J.~W.~Moll, Titus Neupert, and Liang Wu for insightful discussions.  F.~J.~acknowledges funding from Grant PID2021-128760NB0-I00 from the Spanish MCIN/AEI/10.13039/501100011033/FEDER, EU and from a 2024 Leonardo Grant for Scientific Research and Cultural Creation, BBVA Foundation. M.~H.~F. acknowledges support from the Swiss National Science Foundation (SNSF) through Division II (number 207908).
\appendix

\section{Extended symmetry analysis}\label{appA}

The symmetry generators for the CDW order parameters are given as follows. Using $M_n$ and $L_n$ for the components of the $M_1^+$ and $L_2^-$ order parameters, the representation is 
 \begin{align}
 C_3 M_{n} &= M_{n+1}, &  \sigma_h' M_n &= M_n, \nonumber\\  \sigma_v^{(n)} M_n &= M_n, &  \sigma_v^{(n)} M_{n+1} &= M_{n+2},  
\end{align}
and the same applies for $L_n$. Note that the index $n$ is periodic, meaning $M_{n+3} \equiv M_n$. For $U_n$, we have the same transformation under $C_3$ and $\sigma_v$ as above, but crucially
\begin{align}
    \sigma_h' U_n = U_n^*,    
\end{align}
which implies that $\sigma_h'$ (and hence inversion $I= \sigma_h' * C_{2z}$) is broken by a complex $U_n$. 
Finally, translations act on these order parameters as
\begin{align}
t_c M_n = M_n, & & t_c L_n = -L_n, & & t_c U_n = i U_n, 
\end{align}
while in-plane translations act as $t_a = {\rm diag}(1,-1,-1)$, $t_b = {\rm diag}(-1,1,-1)$ for all three OPs.

To connect with notation in previous work, we note Ref. \cite{Wagner23} classified the possible site and bond orders for a monolayer. These are all $\sigma_h$ even irreps but some of them can be $C_{2z}$ odd. The mapping between representations can be seen in table \ref{irreps}. 

\begin{table}[h]
\begin{center}
\begin{tabular}{|c|c|c|c|c|c|c|c|c|} 
 \hline
BCS & $M_1^+$ & $M_1^-$& $M_2^+$& $M_2^-$ & $M_3^+$ & $M_3^-$& $M_4^+$& $M_4^-$ \\ 
 \hline
Ref. \cite{Wagner23}  & $F_1$ &  & $F_2$ & &  & $F_4$ &  & $F_3$ \\ 
 \hline
$\sigma_v$  & $+$ & $-$& $-$& $+$ & $+$ & $-$& $-$& $+$ \\ 
 \hline
$\sigma_h$  & $+$ & $-$& $+$& $-$ & $-$ & $+$& $-$& $+$ \\ 
 \hline
$C_{2z}$  & $+$ & $+$& $+$& $+$ & $-$ & $-$& $-$& $-$ \\ 
 \hline
\end{tabular}
\caption{Monolayer irreps and the mapping to those in Ref. \cite{Wagner23}.}
\label{irreps}
\end{center}
\end{table}
 
\section{Parameter mapping for layer-dependent free energy}
\label{app:coeffs}

The free energy for four layers, $F_4$, can be restricted to the two layer one, $F_2$, by applying the boundary condition $m_n^{l+2} \equiv m_n^l$. Doing this, we find that $F_2=\tfrac{1}{2}(F^{(2)}_2+F^{(3)}_2+F^{(4)}_2)$ with
\begin{align}
F^{(2)}_2  =  \sum_{n,l} (a_1+a_3) (m_n^l)^2 + a_2 m_n^lm_n^{l+1} .
\end{align}
As the free energy is defined per layer, if we set all couplings higher than first neighbor to zero, this free energy is simply equivalent to the two-layer free energy. In general, however, the second neighbor coupling $a_2$ adds up to the intralayer term $a_1$ so an ab initio calculation can only constrain the sum $a_1+a_3$. The same thing occurs for the cubic couplings
\begin{align}
F^{(3)}_2 &= \sum_{l}(b_1+ 6 b_3) m_{1}^{l}m_{2}^{l}m_{3}^{l} \nonumber\\&+ \sum_{n,l}(2 b_2+2b_4) m^l_{n}m^{l}_{n+1}m^{l+1}_{n+2}  
\end{align}
and the quartic ones
\begin{align}
&F^{(4)}_2 = c_1 \sum_l [\sum_n (m_n^l)^2]^2 + \sum_{l,n} \Big[c_2  (m_n^l)^2(m_{n+1}^l)^2 \nonumber \\ &+  c_3 (m_n^l)^2 (m_n^{l+1})^2 + 2c_4 (m_n^l)^3 m_n^{l+1} \nonumber \\
&+ 2c_5 m_n^l m_n^{l+1}\left((m_{n+1}^{l+1})^2+(m_{n+2}^{l+1})^2\right) \nonumber \\ &+c_6 m_n^lm_{n+1}^l m_n^{l+1}m_{n+1}^{l+1} +2c_7  (m_n^l)^2  (m_{n+1}^{l+1})^2  \Big].  
\end{align}

To map these coefficients to those computed ab-initio, we set all second neighbor couplings to zero ($a_3=b_3=b_4=0$). We use the Fourier transforms
\begin{equation}
    \begin{pmatrix} m_n^1 \\ m_n^2 \end{pmatrix} = \begin{pmatrix} 1 & 1 \\ 1& -1 \end{pmatrix} \begin{pmatrix} M_n \\ L_n \end{pmatrix} \label{FTML}
\end{equation}
and
\begin{equation}
    \begin{pmatrix} M_n \\ L_n \end{pmatrix} = \frac 1 2 \begin{pmatrix}1 & 1 \\ 1& -1\end{pmatrix} \begin{pmatrix}m_n^1\\m_n^2\end{pmatrix}
\end{equation}
and compare with the ML free energy in Eqs. \eqref{eq:FM}, \eqref{eq:GLL} and \eqref{eq:FML}. This free energy in momentum space is already an intensive quantity (measured in meV/formula unit cell). Reference~\cite{Ritz23} computed per $2 \times 2 \times 2$ CDW unit cell, and the coefficients were obtained as $a_M=-2.53$, $a_L=-2.92$, $b_M=-22.16$, $b_{ML}=-24.15$, 
$c_M^{(1)}=76.43$, $c_M^{(2)}=-137.67$, $c_L^{(1)}=89.93$, $c_L^{(2)}=-194.47$, $c_{ML}^{(1)}=24.20$, $c_{ML}^{(2)}=347.73$ and $c_{ML}^{(3)}=332.28$ in units of eV/\textup{~\AA}$^{n}$ for the coefficients of order $n$. 

This means to have the free energy per layer we also have to divide it by 1/2, so we match
\begin{align}
\frac{1}{2}(F_M+F_L+F_{ML}) = F_2   
\end{align}
with $F_2$ expressed in terms of $M_n$ and $L_n$ through Eq. \eqref{FTML}. With this procedure, we obtain, for second-order coefficients
\begin{align}
a_1 &= \frac{1}{4}(a_M+a_L) = -1.36\\
a_2 &= \frac{1}{4}(a_M-a_L) = 0.10,
\end{align}
for third-order coefficients
\begin{align}
b_1 &= \frac{1}{8}(b_M+3b_{ML})=-11.83\\
b_2 &= \frac{1}{16}(b_M-b_{ML}) = 0.12,
\end{align}
and for fourth-order coefficients 
\begin{align}
c_1 &= \frac{1}{16} (c_L^{(1)} + c_M^{(1)} + c_{ML}^{(1)} + c_{ML}^{(3)})= 32.68\\ 
c_2 &=\frac{1}{16} (c_L^{(2)} + c_M^{(2)} + c_{ML}^{(2)}  -2c_{ML}^{(3)}) = -40.56\\
c_3 &= \frac{1}{16}(3 c_L^{(1)} + 3 c_M^{(1)} - c_{ML}^{(1)} - c_{ML}^{(3)})=8.91\\
c_4 &= \frac{1}{8} (-c_L^{(1)} + c_M^{(1)})=-1.69\\
c_5 &= \frac{1}{16} (-2 c_L^{(1)} - c_L^{(2)} + 2 c_M^{(1)} + c_M^{(2)})=1.86\\
c_6 &= \frac{1}{8} (2 c_L^{(1)} + c_L^{(2)} + 2 c_M^{(1)} + c_M^{(2)} - 2 c_{ML}^{(1)}) \nonumber\\
&=-5.97\\
c_7 &= \frac{1}{32} (2 c_L^{(1)} + c_L^{(2)} + 2 c_M^{(1)} + c_M^{(2)} + 2 c_{ML}^{(1)} - c_{ML}^{(2)}) \nonumber\\
&=-9.33.
\end{align}
We refer to this set of parameters as the Ritz parameters according to Ref. \cite{Ritz23}. Note for the quadratic terms, which are expected to depend strongly on temperature, we do not take fixed values but consider them as free parameters. 

A more general set of parameters can be obtained by noting that the two-layer $ML$ \emph{ab initio} calculation only provides a constraint on the effective two-layer parameters, i.e. if we rewrite the cubic free energy 

\begin{align}
F^{(3)}_2 &= \sum_{l} b_1^{2L} m_{1}^{l}m_{2}^{l}m_{3}^{l} + \sum_{n,l} b_2^{2L} m^l_{n}m^{l}_{n+1}m^{l+1}_{n+2}  
\end{align}
then
\begin{align}
b_1^{2L} &= b_1+ 6 b_3 =\frac{1}{8}(b_M+3b_{ML})=-11.83\\   
b_2^{2L}/2 &= (b_2+b_4) = \frac{1}{16}(b_M-b_{ML}) = 0.12
\end{align}
and the more general constraint can be written as 
\begin{align}
b_1 &= b_1^{2L} +x\\ 
b_2 &= b_2^{2L}/2 +y\\
b_3 &= -x/6 \\
b_4 &= -y
\end{align}
with arbitrary values of $(x,y)$. Formally the same procedure could be applied to the quartic coefficients as well, but we expect their influence on the phase diagram to be less important than that of the cubic terms.

\section{Parametrization of states} \label{app:parametrization}

In this section we provide the parametrization of $m_n^l$ for all states found in this work. In addition, for ease of comparison with other works, we also provide the parametrization in terms of $MLU$ order parameters following the notation of the commonly used ISOTROPY software. Since in this convention the origin is taken at the A (alkali atom) site, the action of the symmetry elements on $U_n$ is different and hence we will call it $\bar{U}_n$, which is related to $U_n$ by $U_n=e^{i\pi/4}\bar{U}_n$. The relationship between the $ML\bar{U}$ and $m_n^l$ order parameters is 
\begin{align}
m_n^l = M_n + \cos (l \pi) L_n + \Re (e^{i\pi/4} e^{i l \pi/2} \bar{U}_n)    
\end{align}
which means that when $\bar{U}$ is real, this corresponds to a pattern in the V layers as $(1,1,-1,-1)$. The order parameter components are listed as 
\begin{align}
(M_1;M_2;M_3|L_1;L_2;L_3|\Re \bar{U}_1,\Im \bar{U}_1;\Re \bar{U}_2,\Im \bar{U}_2;\Re \bar{U}_3,\Im \bar{U}_3).   \nonumber 
\end{align}
In the real space representation, we simply list the matrix $m_n^l$ with $n=1,2,3$ and $l=1,2,3,4$. With this notation, the parametrization of all the states found in this work is displayed in Table \ref{StatesTable} in both representations along with the space group. The rigid layer states are obtained when all the parameters in the real space representation take the same value $|m_n^i|=m_0$.

\begin{table*}
\begin{center}
\begin{tabular}{|>{\centering\arraybackslash}p{1.4cm}|c|c|>{\centering\arraybackslash}p{1.4cm}|c|c|} 
 \hline
SG & $ML\bar{U}$ & $m_n^i$ & SG & $ML\bar{U}$ & $m_n^i$ \\ 
 \hline
 P6/mmm (A) & $(a;a;a|0;0;0|0,0;0,0;0,0)$ & $\left( \begin{array}{ccc}
    \alpha & \alpha & \alpha   \\
    \alpha & \alpha & \alpha \\
    \alpha & \alpha & \alpha   \\
    \alpha & \alpha & \alpha 
\end{array} \right)$ &  
Fmmm (AABB)  & $(a;0;0|0;0;0|0,0;b,0;b,0)$ & $\left( \begin{array}{ccc}
    \alpha & \beta & \beta   \\
    \alpha & \beta & \beta \\
    \alpha & -\beta & -\beta   \\
    \alpha & -\beta & -\beta 
\end{array} \right)$\\ 
\hline
 P6/mmm (Aa) & $(a;a;a|b;b;b|0,0;0,0;0,0)$ & $\left( \begin{array}{ccc}
    \alpha & \alpha & \alpha   \\
    -\beta & -\beta & -\beta \\
    \alpha & \alpha & \alpha   \\
    -\beta & -\beta & -\beta 
\end{array} \right)$ &  
Cccm (ABAC)  & $(b;b;a|c;c;d|e,e;-e,-e;,0,0)$ & $\left( \begin{array}{ccc}
    \alpha & \alpha & \beta   \\
    \gamma & -\delta & -\epsilon \\
    \alpha & \alpha & \beta   \\
    -\delta & \gamma & -\epsilon 
\end{array} \right)$\\ 
\hline
 P6/mmm (AAAa) & $(a;a;a|b;b;b|c,c;c,c;c,c)$ & $\left( \begin{array}{ccc}
    \alpha & \alpha & \alpha   \\
    \beta & \beta & \beta \\
    \alpha & \alpha & \alpha   \\
    -\gamma & -\gamma & -\gamma 
\end{array} \right)$ &  
Fddd (ABCD)  & $(0;0;0|0;a;0|b,0;0,0;0,b)$ & $\left( \begin{array}{ccc}
    \alpha & \beta &  \alpha   \\
  \alpha &  -\beta &  -\alpha \\
   -\alpha & \beta &  -\alpha   \\
   -\alpha & -\beta &  \alpha 
\end{array} \right)$\\ 
\hline
 Fmmm (AB) & $(a;0;0|0;b;b|0,0;0,0;0,0)$ & $\left( \begin{array}{ccc}
    \alpha & \beta & \beta   \\
    \alpha & -\beta & -\beta \\
    \alpha & \beta & \beta   \\
    \alpha & -\beta & -\beta 
\end{array} \right)$ &  
C222 (AABC)  & $(a;a;b|c;-c;0|d,e;d,-e;f,0)$ & $\left( \begin{array}{ccc}
    \alpha & \beta & \gamma   \\
    \beta & \alpha & \gamma \\
    \delta & -\epsilon & -\zeta   \\
    -\epsilon & \delta & -\zeta
\end{array} \right)$\\ 
\hline
 Cmmm (AAAB)  & $(a;b;b|c;d;d|e,e;f,f;f,f)$ & $\left( \begin{array}{ccc}
    \alpha & \beta & \beta   \\
    \gamma & \delta & \delta \\
    \alpha & \beta & \beta   \\
    \epsilon & -\zeta & -\zeta 
\end{array} \right)$ &  
F222 ($\chi$-ABCD)  & $(0;a;0|0;b;0|c,d;0,0;d,c)$ & $\left( \begin{array}{ccc}
    \alpha & \beta &  \alpha   \\
  \gamma &  -\delta &  -\gamma \\
   -\alpha & \beta &  -\alpha  \\
   -\gamma & -\delta &  \gamma
\end{array} \right)$\\ \hline
\end{tabular}
\caption{Parametrization of all the ground states found in this work. The MLU parametrization follows the ISOTROPY notation. All entries in both MLU and $m_n^i$ are taken positive unless otherwise noted. Each entry corresponds to a particular instance of the manifold of degenerate ground states. Note the state Aaaa is obtained from AAAa by a global sign reversal, the state ABAb is obtained from AAAB by reversing $\delta$, $\epsilon$ and $\zeta$, and the state o-Aaaa is obtained from AAAB by reversing $\alpha$, $\beta$, $\gamma$, $\delta$ and $\zeta$ (which produces aaaA). Hence these states are not listed explicitly in the table.}
\label{StatesTable}
\end{center}
\end{table*}

\section{Additional phase diagrams}\label{app:additional}

\begin{figure}[htbp]
\includegraphics[width=0.35\textwidth]{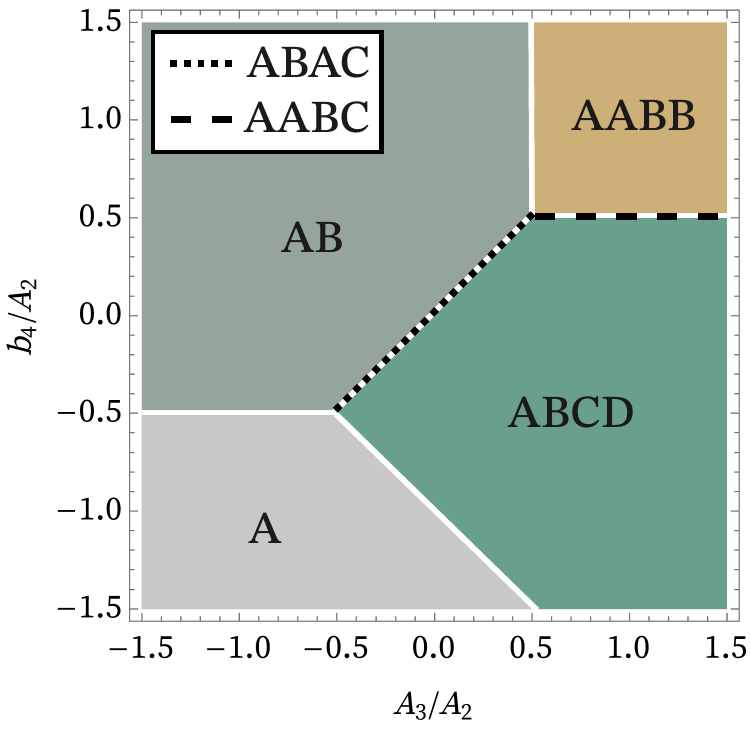}
     \caption{Dimensionless analytical phase diagram of the rigid layer model up to second neighbor couplings with $A_2>0$.}  
     \label{Fig4}
\end{figure}

As shown in the main text, the rigid-layer phase diagram with tri-hex layers only depends on two dimensionless parameters $b_4/A_2$ and $A_3/A_2$. This phase diagram is shown in Fig.~\ref{Fig4}, where the four stable phases are A,AB,ABCD and AABB. In addition, at the phase boundaries between AB and ABCD we find ABAC, and at the boundary between ABCD and AABB we find AABC.  

\begin{figure}[htbp]
\includegraphics[width=0.48\textwidth]{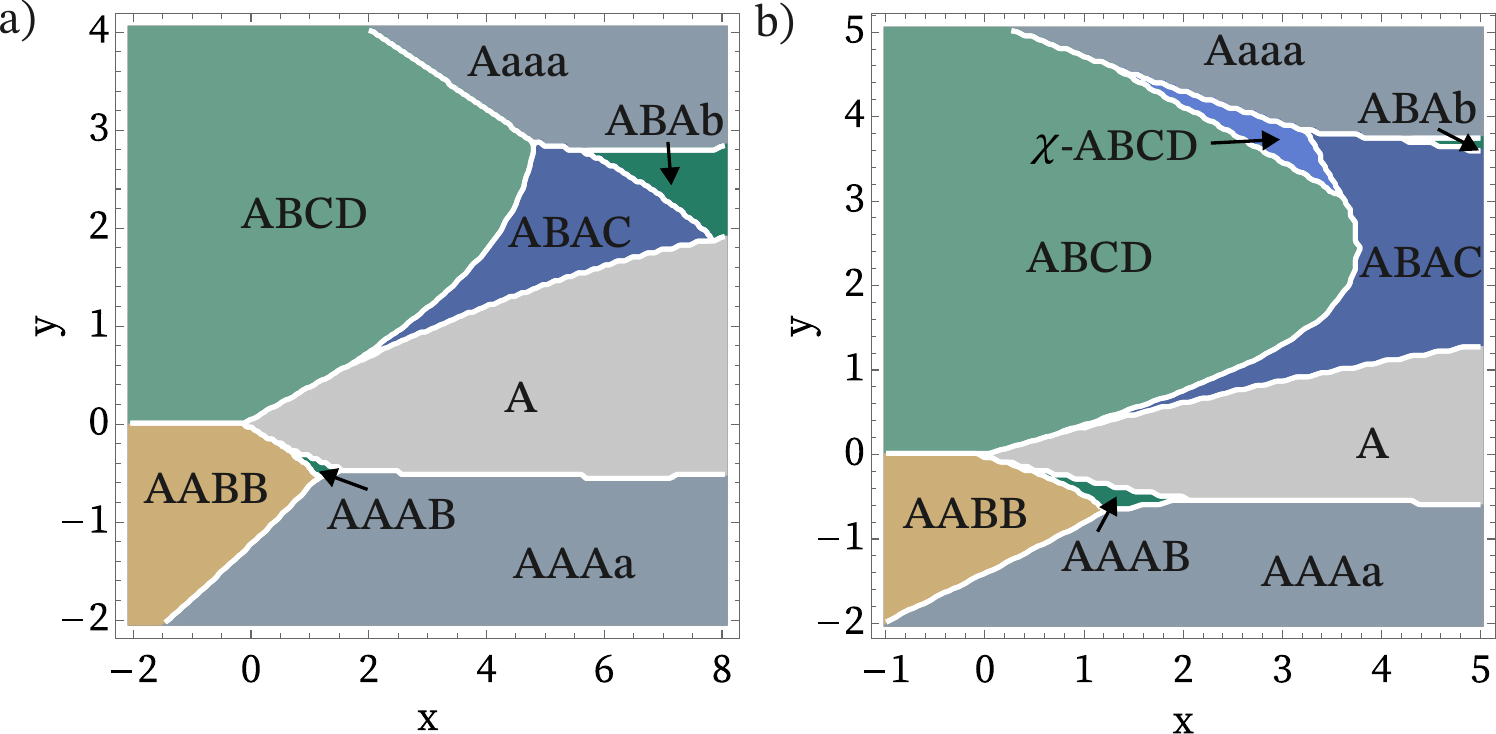}
     \caption{Phase diagrams as a function of the cubic terms in the free energy parametrized by x and y as described in the main text. a) Ritz parameters and b) modified Ritz parameters with $\tilde{c}_2=0.75 c_2$, where the chiral ABCD phase appears. }  
     \label{Fig5}
\end{figure}

In Sec.~\ref{sec:extension}, we described how the four cubic terms in the four-layer model can be chosen to be compatible with the ab-initio results by imposing two constraints, which leave two free degrees of freedom $(x,y)$ to parametrize the resulting manifold. In Fig.~\ref{Fig5}, we show the numerical phase diagrams as a function of these parameters. Figure~\ref{Fig5}(a) shows the plot for ab-initio parameters, which reveals two new phases ABAb (SG, $D_{2h}$) and AAAa (SG $D_{6h}$), but no new chiral phase.  

Figure~\ref{Fig5}(b) shows the same phase diagram with $\tilde{c}_2=0.75c_2$ and $a_2=a_3=0$, where two additional phases emerge at the boundaries between the previous phases: an orthorhombic Aaaa (SG) and a chiral ABCD which is discussed in the main text.

\bibliography{Kagome}
\end{document}